# Structure and electronic transport in graphene wrinkles

Wenjuan Zhu*, Tony Low, Vasili Perebeinos, Ageeth A. Bol, Yu Zhu, Hugen Yan, Jerry Tersoff, and Phaedon Avouris*

IBM Thomas J. Watson Research Center, Yorktown Heights, NY 10598, USA

## Abstract

**Wrinkling is a ubiquitous phenomenon in two-dimensional membranes. In particular, in the large-scale growth of graphene on metallic substrates, high densities of wrinkles are commonly observed. Despite their prevalence and potential impact on large-scale graphene electronics, relatively little is known about their structural morphology and electronic properties. Surveying the graphene landscape using atomic force microscopy, we found that wrinkles reach a certain maximum height before folding over. Calculations of the energetics explain the morphological transition, and indicate that the tall ripples are collapsed into narrow standing wrinkles by van der Waals forces, analogous to large-diameter nanotubes. Quantum transport calculations show that conductance through these *collapsed wrinkle* structures is limited mainly by a density-of-states bottleneck and by interlayer tunneling across the collapsed bilayer region. Also through systematic measurements across large numbers of devices with wide *folded wrinkles*, we find a distinct anisotropy in their electrical resistivity, consistent with our transport simulations. These results highlight the coupling between morphology and electronic properties, which has important practical implications for large-scale high-speed graphene electronics.**

The advance of a new generation of very high speed graphene electronics[1-8] depends upon understanding and controlling the interaction of graphene with the surroundings, especially the supporting substrate[9-11]. Graphene obtained by Chemical Vapor Deposition (CVD) on metals[12-15] is emerging as a powerful platform for graphene electronics with wafer-scale compatible fabrication. However, carrier mobilities in CVD graphene are smaller than for exfoliated graphene[12-15], and the limiting electron scattering mechanisms in these large-scale graphene wafers are not well understood. Recently, several studies have reported on the polycrystalline nature of CVD graphene and the associated electrical resistances due to grain boundaries[16-19], which can be significant. Fortunately, improvements in current state-of-the-art growth techniques have led to large micron scale grain sizes[16-18]. Hence, in typical sub-micron graphene devices, electron scattering mechanisms at length-scale much smaller than the grain size are likely responsible for the inferior carrier mobilities.

A particular issue for CVD graphene is the presence of wrinkles. These are formed by differential thermal expansion, as the metal contracts more than the graphene during post-growth cooling, leaving an excess area of graphene[20-24]. Wrinkles are still present after transfer to the device substrate, as can be clearly seen in SEM images. Figure 1 (a) shows such an image on a $SiO_2$/Si substrate. The dark lines correspond to the graphene wrinkles, since these regions reflect fewer secondary electrons. The same features are also reflected in the Raman images. Figure 1 (b) shows the Raman G band intensity map of the same graphene area shown in Fig. 1 (a). It displays the total intensity of the G phonon band. There are clear line features in the phonon spectrum at the same locations as the features in the SEM image. The total G band intensity is higher at the wrinkles due to the extra layers of graphene as compared to the surrounding area[25]. In Fig. 1 (c) we show a map of the ratio of the Raman D to G band intensities. We note that there is no corresponding line feature in the D to G ratio map at the position of the wrinkle, indicating that this feature is not a line defect, such as grain boundary that gives rise to an increased D band[18,26]. A TEM cross-sectional image of such a wide graphene wrinkle is shown in Fig. 1 (d). The width of this wrinkle is about 50nm. The profile of TEM intensity in the boxed area is shown in the inset. The multi-layer structure of the wrinkle clearly shows that it is not a standing wrinkle, but a folded structure[20].

More quantitative information about the wrinkle distribution is obtained using AFM with "SuperSharp" tips. Figures 2 (a-b) show the AFM images of a wide graphene wrinkle (~135nm) and a narrow graphene wrinkle (~16nm). The AFM heights of the marked region in (a) and (b) are shown in Fig. 2 (c) and are about 0.9 nm and 3.8 nm, respectively. Figure 2(d) gives the statistical distribution of the wrinkle heights as a function of their width. The wide wrinkles all have similar height, roughly 1 nm, but they exhibit a broad range of widths. In contrast, the narrow wrinkles are taller and have a broad distribution of heights, ranging roughly from 2 to 6 nm. Their width is apparently below the resolution of our AFM.

Little is known about the factors determining the distribution of wrinkles, which may depend on the details of CVD growth and subsequent transfer. However, previous studies have already yielded insight into the morphology of individual wrinkles. Figure 2(e-g) shows three distinct morphologies: 2(e) is the simple ripple geometry usually assumed[23,27-31]; 2(g) is the recently identified folded geometry[20]; and we propose an intermediate geometry, 2(f), a *standing collapsed wrinkle*. The sequence of shape is easily understood at the qualitative level, as arising from the competition between elastic bending and van der Waals binding [20,32]. The ripple structure minimizes the bending distortion. However, when there is a large amount of excess material, collapsing it to form bilayers or folding it to form trilayers provides additional van der Waals binding, at the cost of progressively increased bending. For a given structure, the binding increases with length, while the bending energy is constant (roughly proportional to the number of sharp bends), leading to the sequence of shapes.

To better understand the energetic competition and resulting morphologies, we perform total energy calculations, including the van der Waals binding and elastic bending and assuming a variational form for the geometry. The only material parameters are the bending stiffness κ and the van der Waals binding energy β (for details see Supplemental information). We use the value $\kappa$=1.4 eV [33,34] for the bending stiffness of graphene, and $\beta$=40 meV per carbon atom[32] for the van der Waals adhesion energy between the graphene sidewalls. When the excess graphene length (i.e. the graphene length compared to the substrate length) is greater than $L_m \approx 24\sqrt{\kappa/2\beta} \approx 16.3$ nm, we find that the *folded* structure [Fig. 2(g)] has lowest energy. Below this, the *standing collapsed wrinkle* [Fig. 2(f)] has lower energy. The height of the *standing wrinkle* at the transition (i.e. the maximum height) is about 8.4 nm, very close to $L_m/2$. Taking into account the uncertainties in AFM of flexible structures and in the parameters κ and β, this estimate is consistent with the experimental value found above of 5-6 nm.

For electronic devices, the most important issue for graphene wrinkles is their impact on transport. Here we address this both experimentally and theoretically. For *folded wrinkles*, we systematically measured the electrical resistivities both along the fold and across it, using four-probe devices as illustrated in Fig. 3 (a-b). Their averaged resistances, herein denoted as ⟨$R_{al}$⟩ and ⟨$R_{ac}$⟩ respectively, are plotted in Fig. 3 (c-d) as functions of the gate voltage $V_g$ and compared with resistances of the control (flat) devices ⟨$R_0$⟩. The resistances involving the fold structure exhibit an interesting anisotropy: **(i)** ⟨$R_{al}$⟩ is smaller than ⟨$R_0$⟩, especially when biased near the charge neutrality point. **(ii)** ⟨$R_{ac}$⟩ shows no difference from ⟨$R_0$⟩ within error. We cannot directly measure the resistance of *standing wrinkles*, because their high density precludes contacting them individually. Hence, each device with the *folded wrinkles* has a control device constructed beside it, which measures the resistances associated with these smaller background wrinkles. $\langle R_{al} \rangle$, $\langle R_{ac} \rangle$ and $\langle R_0 \rangle$ are obtained by averaging over dozens of these devices.

In general, folds could trap impurities that act as dopants or scatterers, affecting the resistance. Such extrinsic effects would depend on the details of processing. Our measurements suggest that trapped impurities are not a significant factor here, see Supp. Info. We therefore address theoretically the intrinsic resistance of folds, without any such impurities or defects. Electrical transport *along folded wrinkles* can be analyzed using a simple diffusive transport model. Roughly speaking, the *folded wrinkle* can be viewed as a strip of trilayer graphene. Neglecting hybridization between layers, the electrostatic problem for the tri-layer can be solved self-consistently. The resulting carrier distributions within each layer are plotted in Fig. 4(a). Due to non-linear charge screening[35], the carriers are almost all confined to the bottom layer for large $V_g$, while the carriers are more equally distributed when the device is biased near the charge neutrality point. The transport coefficient (i.e. the mobility) is commonly seen to improve with decreasing carrier density[10,36,37], and we confirm this behavior in our own devices (see Suppl. Info. Sec. 6). Therefore the charge redistribution in the tri-layer structure should improve its effective carrier mobility relative to that of monolayer graphene at a given electrostatic doping. We confirm this by quantitatively modeling the experiments as shown in Fig. 4(b) (see Suppl. Info. Sec. 3). The largest difference in electrical resistance between $\langle R_{al} \rangle$ and $\langle R_0 \rangle$ occurs at the charge neutrality point, when charge redistribution in the tri-layer graphene is most effective.

Electrical transport *across folded wrinkles*, however, cannot be explained with the simple diffusive model above. The excess length associated with the fold and its reduced doping would both increase the resistance. If this were the controlling physics, the resistance would be an order larger than what is experimentally observed, see Fig. 4(c). This discrepancy can be reconciled by taking into account an additional conduction pathway via interlayer tunneling[38,39] across the collapsed bilayer graphene, which can reduce the larger resistance associated with the density-of-states bottleneck.

To estimate the role of inter-layer tunneling in limiting the resistance, we performed quantum transport simulations based on the non-equilibrium Green's function method[40]. Since we are also interested in the resistance of *standing wrinkles*, we perform calculations for the simpler geometry of Fig. 2(f), but for varying excess graphene lengths spanning the range from standing to folded geometries. For both geometries, we expect reduced doping in the raised graphene areas, reflecting the poor coupling to the gate due to its geometry for the *standing wrinkle* or due to screening by the bottom graphene layer for the *folded wrinkle*. For simplicity, we treat these regions as undoped, giving an upper bound on the resistance. The resulting room temperature resistances are shown in Fig. 5(a) as a function of the length $\lambda$ of the collapsed bilayer region. Depending on the orientation of the wrinkle, the bilayer could have some rotated alignment, but we focus on the simpler case with zero misorientation, where the stacking depends only on sliding of one layer over the other. We show the average resistance and range for different bilayer alignment.

The most striking feature of Fig. 5(a) is that the resistance depends only very weakly on the excess graphene length. The reason is seen in Fig. 5 (b), which shows that much of the current flows between graphene layers at the base of the wrinkle, rather than flowing through the whole length. To verify this interpretation, we repeat the calculations with the graphene sheet cut at the top of the wrinkle. Despite totally blocking the direct intralayer pathway, the change in resistance is minor. We expect the interlayer transport to be similar for *folded wrinkles*, since it occurs largely near the base. Thus, our calculations suggest a resistance on the order of $\approx 200\Omega\cdot\mu m$, which is relatively independent of $\lambda$, for either type of wrinkles. The resistance observed in our experiments is also very small, below the experimental accuracy, consistent with the small calculated value.

In conclusion, we reported experimental and theoretical measurements of collapsed graphene wrinkles, and compared these. The calculated energetics based on competition between the elastic and van der Waals energies is consistent with the experimental observation of a maximum wrinkle height of ~6nm, substantiating our physical picture of the structure of *standing* and *folded wrinkles*. Our transport experiments on these *folded wrinkles* yield a distinctive anisotropy in the fold resistivity consistent with our model. We conclude that this anisotropy arises because transport along and across the *folded wrinkles* are limited by different transport effects: diffusive transport of the charge distributed across the multi-layered folds, vs local interlayer tunneling across the collapsed region. From an applications standpoint, the former degrades the on-off ratio, while the latter can contribute a significant resistance to the overall device, of the same order as typical graphene contact resistances[41,42]. Our study therefore identifies a source of electrical performance degradation in CVD grown graphene, and highlights the subtle interplay between morphology and electronic properties. It also underscores the importance of a better fundamental understanding of the formation and engineering of wrinkles.

## Methods

The method we used in this work to prepare monolayers of graphene is based on Chemical Vapor Deposition (CVD) of graphene on Cu and is similar to the method described in reference 13. A Cu foil (25 μm thick, 99.98%, Alfa-Aesar) was placed in a 1–inch diameter quartz furnace tube at low pressure ($10^{-6}$ Torr). After evacuation, the Cu foil was heated to $1050^{o}$C in vacuum. At this temperature the sample was exposed to ethylene (6 sccm, 6 mTorr) for 10 minutes. The sample was then cooled down under vacuum. PMMA resist was spin-coated on top of the graphene layer formed on one side of the Cu foil. The Cu foil was then dissolved in 1 M iron chloride. The remaining graphene/PMMA layer was washed with DI water, 1M HCl and DI water and transferred to the Si/$SiO_2$ substrate. Subsequently, the PMMA was dissolved in hot acetone ($80^{o}$C) for one hour. The substrate with the transferred graphene was then rinsed with methanol and dried in a stream of nitrogen.

Subsequently, metal alignment marks were formed by lift-off and graphene Hall-bar structures were fabricated by photoresist patterning and $O_2$ plasma etching. Following that, SEM, AFM and Raman characterization was performed, and then source/drain and sensing terminals were formed using Ti/Pd/Au metallization and lift-off. The SEM measurements were obtained at 3KV. The Raman G band intensity is summed in the frequency range of 1480 $cm^{-1}$ to 1700 $cm^{-1}$ and the D band intensity is summed in the frequency range of 1253 $cm^{-1}$ to 1450 $cm^{-1}$. The AFM images were taken in the tapping mode with "SuperSharp" silicon tip. The tip radius is less than 5nm. For TEM analysis, the sample was prepared using dual-beam focus ion beam, and imaged in a JOEL 3000F TEM operated at 300eV.

## Acknowledgement

We would like to thank B. Ek, J. Bucchignano, and G. P. Wright for their contributions to device fabrication. We would also like to thank X. Li, M. Freitag, F. Xia, Y. Wu, D. Farmer, Y.-M. Lin, G. Tulevski and C. Y. Sung for their insightful discussion and help in the project. T. Low gratefully acknowledges use of a computing cluster provided for by Network for Computational Nanotechnology, and partial funding from INDEX-NRI. The authors would also like to thank DARPA for partial financial support through the CERA program.

**Figure Captions:**

**Figure 1. Physical characterization of CVD graphene on SiO$_2$/Si substrate. (a)** SEM image of graphene on a SiO$_2$/Si substrate. **(b)** Raman G band intensity map of graphene on SiO$_2$/Si substrate. **(c)** Map of the D to G Raman band intensities ratio of graphene on SiO$_2$/Si substrate. **(e)** TEM cross-section of graphene on SiO$_2$/Si substrate. On top of the graphene are PMMA resist and Cr layers. The inset shows the profile of the TEM contrast in the red rectangle area.

**Figure 2. Wrinkle topographies and structures. (a)** AFM image of graphene on SiO$_2$/Si substrate. The field of view is 3 μm. **(b)** AFM image of a narrow wrinkle. Field of view is 100 nm. **(c)** The step profile of a wide and a narrow wrinkle. **(d)** Statistical distribution of wrinkle heights as a function of their width. Schematic illustration of three classes of graphene wrinkles: **(e)** simple *ripple*; **(f)** *standing collapsed wrinkle*; **(g)** *folded wrinkle*. For details on these geometries see the Supplemental Information.

**Figure. 3: Transport measurements along/across *folded wrinkles*. (a-b)** Device layout of the Hall-bars structures across and along the fold. Control devices without fold are fabricated alongside. **(c)** Averaged resistances $\langle R_{al} \rangle$ as a function of gate voltage $V_g$ for devices along the fold, compared with control devices $\langle R_0 \rangle$. Statistical averages are obtained using the geometric mean sampled over 42 devices. Dashed lines indicate the statistical standard deviations. **(d)** Same as (c), but for devices measured across the fold $\langle R_{ac} \rangle$. Device length is 2 μm and width is about 0.35 μm (along the fold) and 0.3 μm (across the fold). The fold's width in these devices was 0.14 μm.

**Figure. 4: Electrostatics and diffusive transport in graphene fold. (a)** Electrostatic modeling of the carrier distribution in a tri-layer graphene system, $n_{1,2,3}$, as function of $V_g$. $n_1$ is the layer closest to the gate. Finite electron-hole puddles densities of $6.5 \times 10^{11} cm^{-2}$ is deduced from Hall measurement. **(b-c)** Modeling of resistance (b) $\langle R_{al} \rangle$ along the fold, and (c) $\langle R_{ac} \rangle$ across the fold, as function of $V_g$, using a diffusive model as described in the main text (see also Suppl. Info.). The diffusive model works well for $\langle R_{al} \rangle$ but fails to capture the behavior of $\langle R_{ac} \rangle$.

**Figure. 5: Quantum transport modeling across a *collapsed wrinkle*. (a)** Quantum transport modeling of the room-temperature resistance of the *standing collapsed* graphene wrinkle as function of the length of the collapsed bilayer λ, averaged over 20

samples with different bilayer stacking alignment (Solid square symbol, with standard deviations indicated). For comparison, open circles show similar calculations where we cut the top of the wrinkle to suppress purely intra-layer current. We use the non-equilibrium Green's function (NEGF) method. Details of the method and calculations are described in Supp. Info. **(b)** Atomic structure and interlayer tunneling for a *standing collapsed* graphene wrinkle. Bubble plot shows the out-of-plane current density, with the bubbles' radius being proportional to the magnitude of current density.

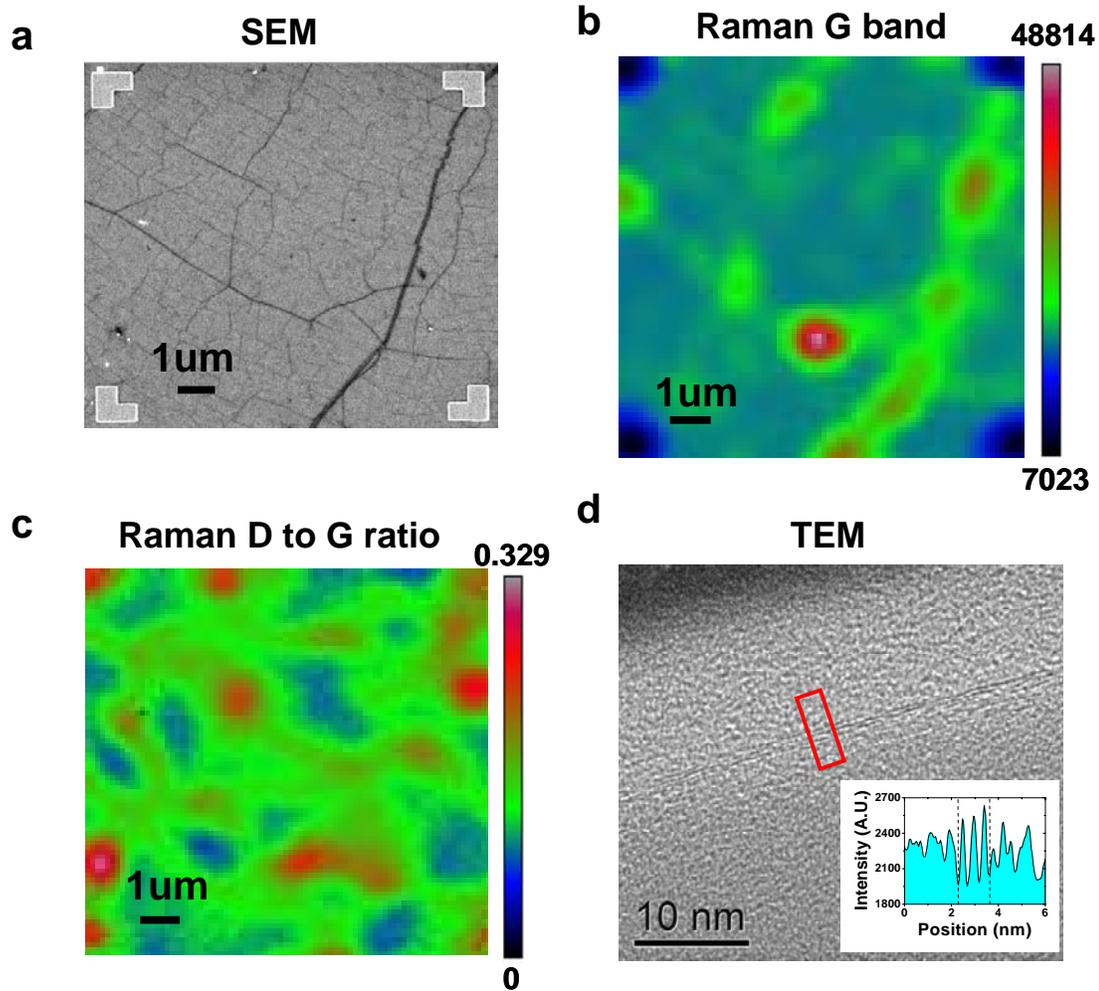

**Figure 1**

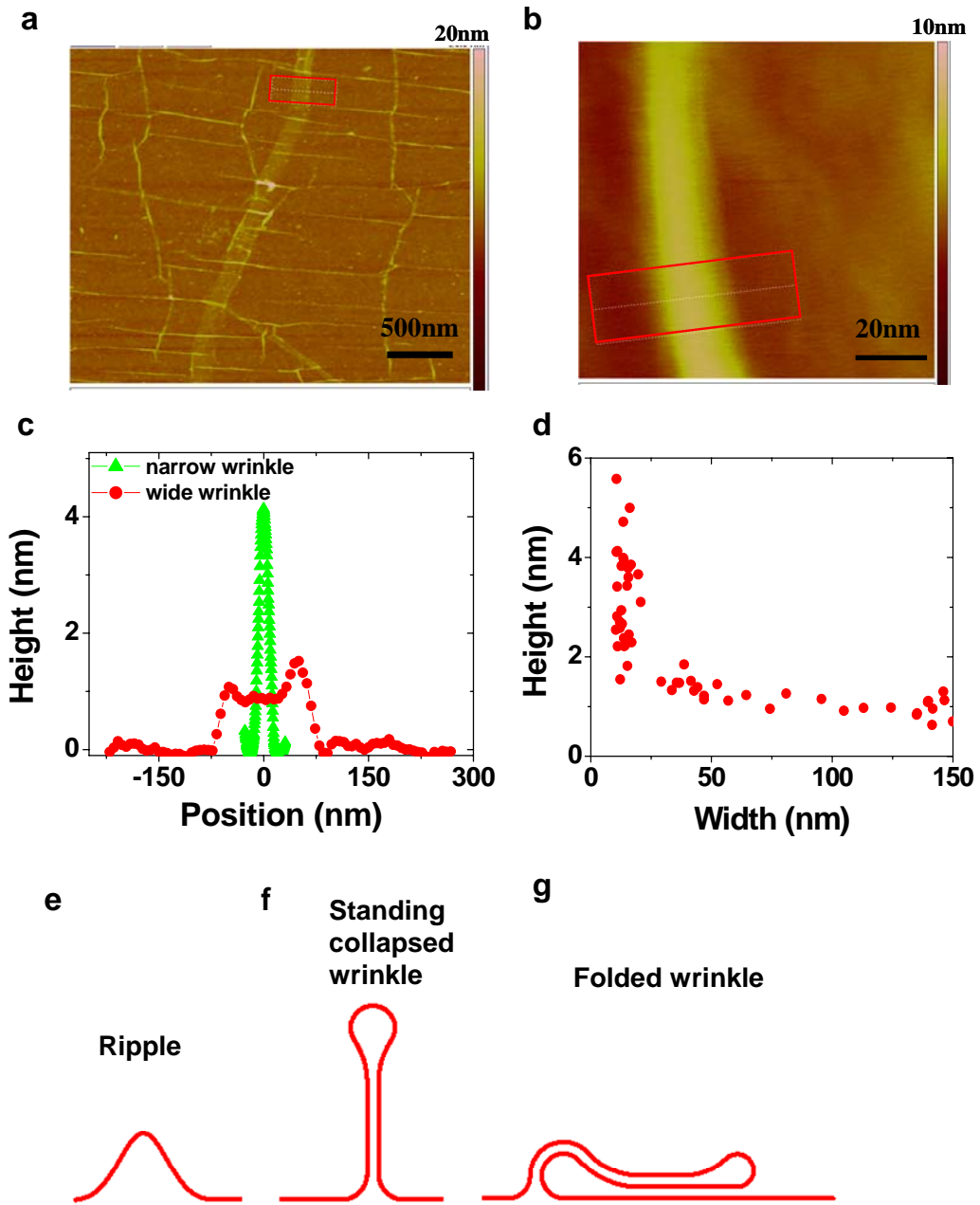

**Figure 2**

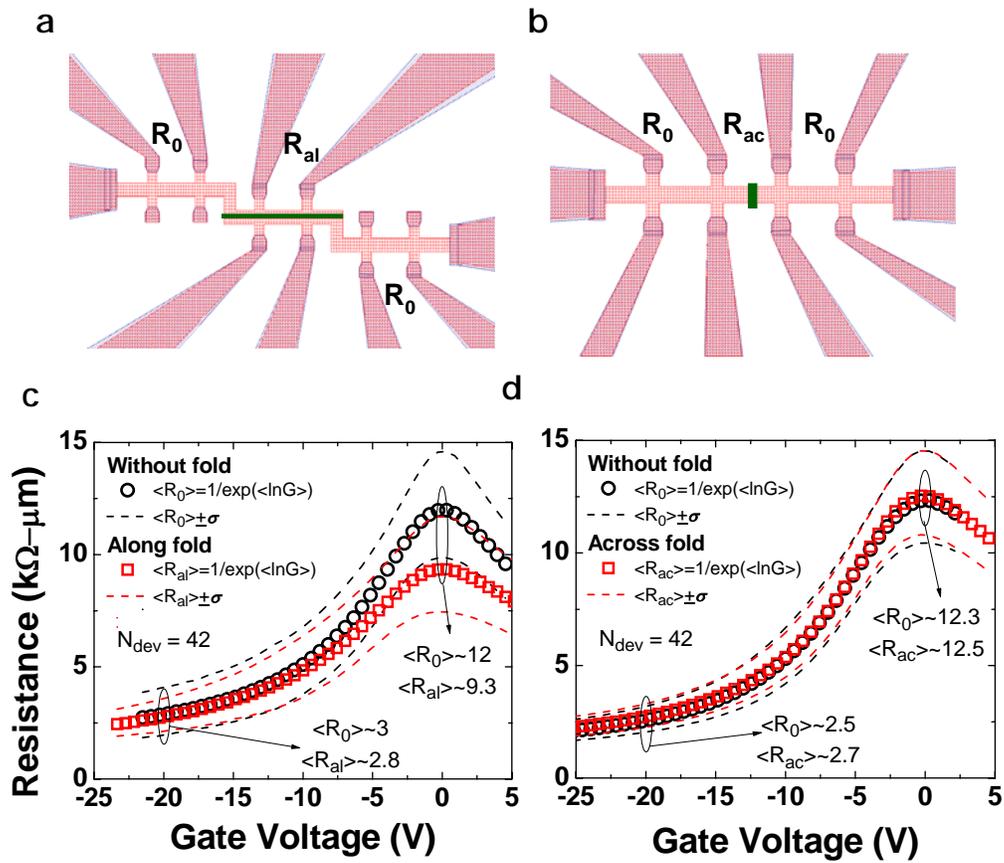

**Figure 3**

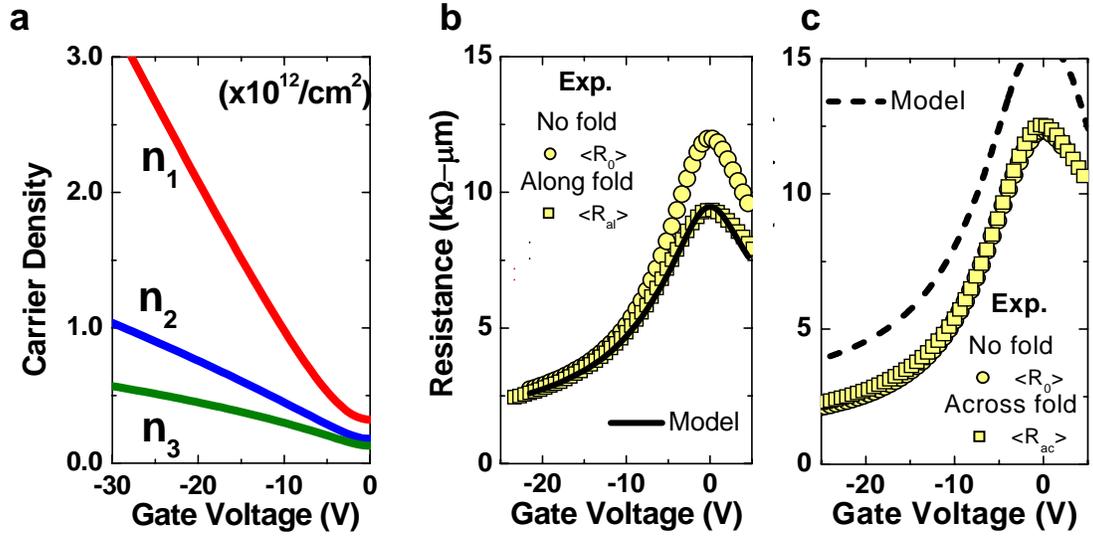

**Figure 4**

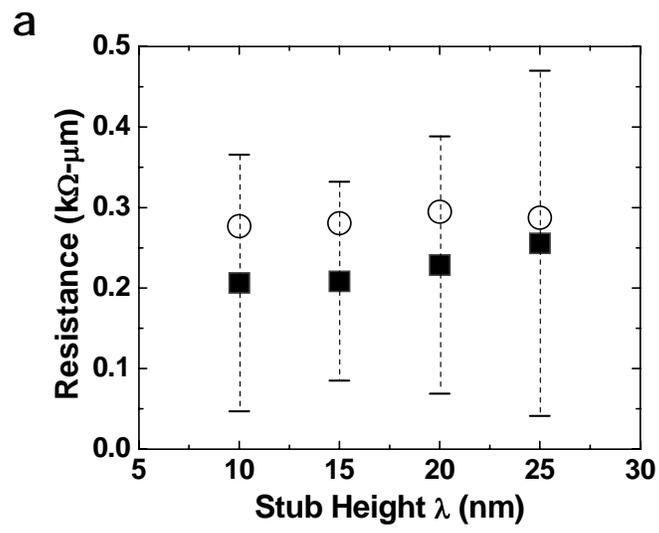

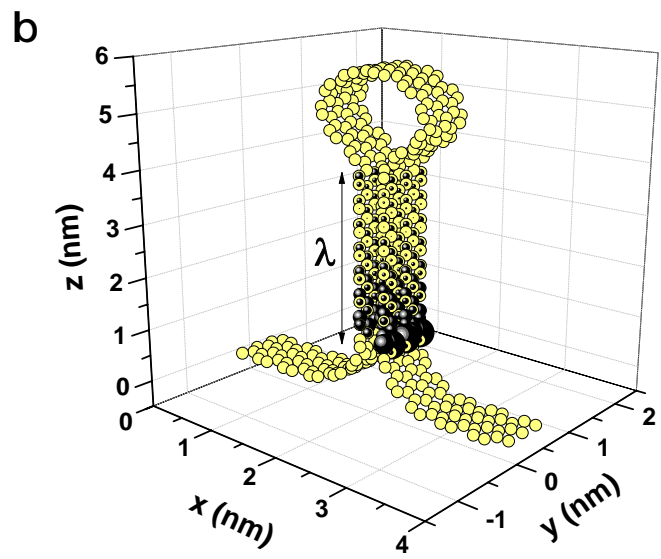

**Figure 5**

# Supplementary Information:
# Structure and electronic transport in graphene wrinkles

Wenjuan Zhu*, Tony Low, Vasili Perebeinos, Ageeth A. Bol, Yu Zhu, Hugen Yan,
Jerry Tersoff and Phaedon Avouris*

IBM Thomas J. Watson Research Center, Yorktown Heights, NY 10598, USA

## 1. Maximum height estimate of the *standing collapsed wrinkle*.

To estimate the maximum height, we assume that the wrinkle has a fixed amount of material (i.e. no sliding of graphene along the surface), and that it adopts the minimum-energy morphology. We therefore need to compare energies of the different structures for a fixed amount of excess graphene in the wrinkle. First, we estimate energy of the *folded wrinkle*. As illustrated in Fig. S1, the structure of *folded wrinkle* consists of the right and left bulb-shaped curves with similar radiuses and a flat trilayer region. The right bulb we approximate by a pair of arcs, concave and convex. The two left bulbs are approximated by the arcs of the same angles, with radiuses being different by the van der Waals distance $h$ separating graphene layers. The trilayer region has length $\lambda$, and the bilayer has length $\lambda+\xi$. The base is approximated by arcs of angle $\pi/2$ and radius $R_b$. The energy of the *folded wrinkle* in this model is given by:

$$E_f = \frac{\kappa}{2}\left(\frac{\pi}{R_b} + \frac{\theta_1}{R_1} + \frac{\theta_1}{R_1+h} + \frac{\theta_2}{R_2} + \frac{\theta_2}{R_2+h} + \frac{\theta_3}{R_3} + \frac{\theta_4}{R_4}\right)$$
$$-\beta\left(\theta_1\left(R_1+\frac{h}{2}\right) + \theta_2\left(R_2+\frac{h}{2}\right) + 2\lambda + \xi\right) \qquad (1)$$
$$+\beta_{sub}(2R_b + h)$$

where $\kappa$ is graphene bending stiffness and $\beta$ is van der Waals adhesion energy. The first term reflects the bending energy, the second term reflects adhesion energies of the bilayer and trilayer regions, and the last term reflects the adhesion energy cost to peel off graphene from the substrate. We will use $\beta_{sub}=\beta$. The excess length is defined as the length of the graphene fold minus the length of the flat substrate and it is given by:

$$L = \pi R_b + \theta_1(2R_1+h) + \theta_2(2R_2+h) + 2\lambda + \xi + \theta_3 R_3 + \theta_4 R_4 - (2R_b+h) \quad (2)$$

where relationships between the angles and radiuses are determined by the geometric constraints, see Fig. S1: $\xi = R_4 \sin\theta_4 - R_3 \sin\theta_3$, $\theta_1 - \theta_2 = \pi/2$, $R_1 \sin\theta_1 + R_b = h + (R_2+h)(1-\cos\theta_2)$, $\theta_3 - \theta_4 = \pi$, $R_3(1-\cos\theta_3) = h + R_4(1-\cos\theta_4)$. Minimization of energy in Eq. (1) with respect to the five variational parameters $R_b$, $R_1$, $\theta_1$, $R_3$, $\theta_3$ for a fixed excess length $L$ from Eq. (2) gives the energy of the fold as a function of $L$.

Similarly, we can estimate the energy of the *standing collapsed wrinkle* geometry, shown in Fig. S2:

$$E_{sc} = \frac{\kappa}{2}\left(\frac{\pi}{R_b} + \frac{2\theta_1}{R_1} + \frac{2\theta_2}{R_2}\right) - \beta\lambda + \beta_{sub}(2R_b+h) \quad (3)$$

The excess length here is given by:

$$L = \pi R_b + 2\lambda + 2\theta_1 R_1 + 2\theta_2 R_2 - 2R_b - h \quad (4)$$

where $\theta_1 = \theta_2 + \pi/2$, $R_1 \sin\theta_1 = h/2 + R_2(1-\cos\theta_2)$ are found from the geometrical constraints. Therefore, there are three variational parameters: $R_b$, $R_1$, $\theta_1$ which minimize the energy in Eq. (3).

Commonly used values of $\kappa$=1.4 eV[1] and $\beta$ corresponding to 40 meV adhesion energy per carbon atom[2,3] suggest an intrinsic length scale $R_0 = \sqrt{\kappa/2\beta} \approx 6.8$ Å. Numerical energy minimization from Eq. (1) and (3) using parameters $R_0$=6.8 Å and

$h=3.4$ Å leads to the values of the variational radiuses of the left and right bulbs in the *folded wrinkle*: $R_1 \approx 6.5$ Å, $R_3 \approx 4.9$ Å (see Fig. S1 caption) to be in very good agreement with the values found from the DFT optimized geometry of $5-6$ Å[4]. The minimum energy of the *standing collapsed wrinkle* from Eq. (3) as a function of $L$ is given as $\frac{E_{sc}}{\beta R_0} \approx 14.78 - \frac{L}{2R_0}$, while minimum energy of the *folded wrinkle* from Eq. (1) is given by $\frac{E_f}{\beta R_0} \approx 27.12 - \frac{L}{R_0}$. The equal energy condition $E_{sc} = E_f$ is satisfied for $L_m \approx 24.7 R_0$, which defines a transition height from *standing collapsed wrinkle* to *folded wrinkle* as $R_b + \lambda + R_1(1-\cos\theta_1) + R_2 \sin\theta_2$, where $\lambda \approx 7.9 R_0$ is found from Eq. (4). The height of the *standing wrinkle* at the transition (i.e. the maximum height) is about $12.4 R_0 \approx 8.4$ nm, very close to $L_m/2$.

## 2. Electrostatic modeling of the trilayered folds regions

We model the electrostatics of the graphene fold as a tri-layer graphene system, assuming that the graphene layers are electrically decoupled from one another. Through the Poisson equation, the Dirac point potential in each layer with respect to Fermi energy can be computed as follows,

$$\begin{aligned} V_2 &= -\tfrac{d_0}{\varepsilon_0} n_3 + V_3 \\ V_1 &= -\tfrac{d_0}{\varepsilon_0}(n_2 + n_3) + V_2 \\ V_g &= -\tfrac{1}{C_g}(n_1 + n_2 + n_3) + V_1 \end{aligned} \quad (5)$$

where $V_3$ is given a priori. $d_0 = 3.4$Å is the graphene interlayer separation, $\varepsilon_0$ is the free space permittivity, $C_g$ is the back gate capacitance and $V_g$ is the applied gate bias. In Fig. 4a of the main manuscript, the calculated carrier densities assumed a finite electron-hole puddle densities $n_0 = 6.5 \times 10^{11}$ cm$^{-2}$ estimated from Hall measurements. The fractional carrier population in the graphene layer closest to the gate, i.e. $n_1/n$ where $n = n_1+n_2+n_3$, is closer to unity at larger $V_g$. On the other hand, the layer densities are more equally distributed when $V_g$ is biased near the Dirac point. This carrier redistribution within the trilayered graphene system is a consequence of nonlinear screening[4], and is crucial to explaining our experimental observations.

### 3. Diffusive transport modeling along/across a graphene fold

We discuss first electronic transport *along* a graphene fold. The effective electrical conductivity $\sigma_{eff}$ in the diffusive limit can be written as,

$$\sigma_{eff} = \frac{W_f}{W}(\sigma_1 + \sigma_2 + \sigma_3) + \frac{W - W_f}{W}\sigma \tag{6}$$

where $\sigma_j$ refers to the electrical conductivity in the $j^{th}$ layer and $\sigma$ is the electrical conductivity in monolayer graphene i.e. control devices. $W_f$ is the width of the graphene fold, estimated from SEM to be $\approx 0.14 \mu m$, and $W$ is the device width. In addition, the electrical conductivity $\sigma$ as a function of the carrier density $n$ can be determined through Hall measurements. The carrier mean-free-path, $\lambda_{MFP}(n)$, can simply be derived from $\sigma = \frac{4e^2}{\pi h}\sqrt{\pi n}\lambda_{MFP}$ [5]. If each graphene layer in the fold also follows the same $\lambda_{MFP}(n)$ functional relationship, then the respective $\sigma_j$ are also known. In this case, the calculated $\sigma_{eff}$ is shown in Fig. 4c of the main manuscript, yielding good agreement. Electronic transport *across* a graphene fold can be modeled in similar fashion, with $\sigma_{eff}$ written as,

$$\frac{1}{\sigma_{eff}} = \frac{L_f}{L}\left(\frac{1}{\sigma_1} + \frac{1}{\sigma_2} + \frac{1}{\sigma_3}\right) + \frac{L - L_f}{L}\frac{1}{\sigma_0} \tag{7}$$

where $L_f$ is the length of the graphene fold, estimated from SEM to be $\approx 0.14 \mu m$, and $L$ is the device length.

### 4. Quantum transport modeling of *standing collapsed wrinkle*

Here we elaborate on the electronic transport calculation of the *standing collapsed* graphene wrinkle in the main manuscript. We assume that the transport direction is along the armchair direction, as illustrated in Fig. 5a. The Hamiltonian $H$ is described by a nearest neighbor $p_z$ tight-binding model[6] including both in-plane and out-of-plane couplings,

$$H = \sum_i V_i a_i^\dagger a_i + \sum_{<ij>} t_{ij} a_i^\dagger a_j + \sum_{ij} s_{ij} a_i^\dagger a_j \qquad (8)$$

where $V_i$ denote the on-site energy, $t_{ij}$ the in-plane coupling and $s_{ij}$ the out-of-plane coupling. Explicitly, they are expressed as,

$$t_{ij} = \frac{v_{ij} \times p_i}{r_{ij}} \frac{v_{ij} \times p_j}{r_{ij}} \frac{(v_{ij} \times p_i) \cdot (v_{ij} \times p_j)}{|v_{ij} \times p_i||v_{ij} \times p_j|} \varepsilon_{pp}^\pi$$

$$s_{ij} = -\alpha\gamma \exp\left(-\frac{r_{ij}-r_p}{\delta}\right) \frac{p_i \cdot v_{ij}}{r_{ij}} \frac{p_j \cdot v_{ij}}{r_{ij}} \qquad (9)$$

where $p_i$ refers to the local out-of-plane vector, $v_{ij}$ is the bond vector and $r_{ij} = |v_{ij}|$. Parameters $r_p \approx 0.34 nm$ refers to the equilibrium graphene interlayer separation, $\gamma \approx 0.119 \varepsilon_{pp}^\pi$ is the out-of-plane coupling energy, $\delta \approx 0.185 \times \sqrt{3} r_0$ where $r_0 = 0.142 nm$ is the carbon-carbon bond-length and $\alpha \approx 1.4$ is a fitting parameter [6].

Electronic transport across the structure is calculated using the non-equilibrium green function method[5] within the Landauer formalism, assuming periodic boundary condition along the transverse width direction. The transmission function $T(k_y, E)$ can then be calculated. The finite temperature device conductance can be calculated using,

$$G = \frac{2e^2}{h} \int_{-\infty}^{\infty} \frac{1}{k_B T} \exp\left[\frac{E-\mu}{k_B T}\right] f^2(E) \left[\sum_{k_y} T(k_y, E)\right] dE \qquad (10)$$

where $f(E)$ is the Fermi Dirac distribution. The resistance associated with the *standing collapsed wrinkle* can then be calculated after subtracting off the quantum contact resistance. In our calculations, we assume that the electrostatic doping of the flat region to be 0.2eV, and undoped in regions which are raised, namely the collapsed bilayer and the structure subtended from it. Temperature is taken to be 300K as per experiments.

## 5. Temperature dependence of conductivity

The conductivities of the graphene device at 4.2K and 300K are shown in Fig. S3. We observe that the conductivity is nearly unchanged when the temperature is decreased from 300K to 4.2K.

## 6. Measurements of Hall Mobility

We performed standard Hall measurement to obtain the resistivity tensor and then the conductivities $\sigma_{xx}$ and $\sigma_{xy}$, from which the carrier mobility $\mu$ and carrier density $n$ can be extracted.

$$\mu = \frac{1}{B}\frac{\sigma_{xy}}{\sigma_{xx}} \tag{11}$$

$$n = \frac{\sigma_{xx}(1+\mu^2 B^2)}{\mu q} \tag{12}$$

where B is the magnetic field. These quantities are plotted in Fig. S4, obtained at 300K. We emphasize that the extraction method breaks down when the graphene is biased near the Dirac point, the range highlighted in the plot. The observed downturn in the mobility is unphysical, an artifact of the extraction method which ignores the two carrier nature of transport near the Dirac point [7]. Outside this region, the measured mobility clearly shows a decreasing mobility with increasing doping.

## 7. Dirac point shifts due to folds

The statistical sampling of Dirac voltage for the graphene Hall-bars with and without fold is shown in Fig.S5: (a) across fold vs no fold; (b) along the fold vs no fold. The statistics of Dirac voltage value indicate that the presence of a fold does not lead to significant changes in the Dirac point shifts, hence of the doping level. This indicates that most of the trapped impurities reside in the substrate or the $SiO_2$-graphene interface.

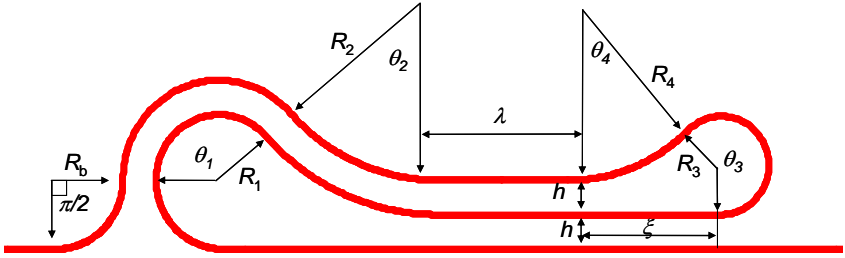

Fig. S1. Schematics of the *folded graphene wrinkle*. The minimum energy from Eq. (1) corresponds to the values of the variational parameters $R_b = R_0$, $R_1 \approx 0.950 R_0$, $R_2 \approx 3.025 R_0$, $\theta_1 \approx 0.764\pi$, $\theta_2 \approx 0.264\pi$, $R_3 \approx 0.714 R_0$, $R_4 \approx 2.555 R_0$, $\theta_3 \approx 1.246\pi$, $\theta_4 \approx 0.246\pi$, where $R_0 = 6.8$ Å.

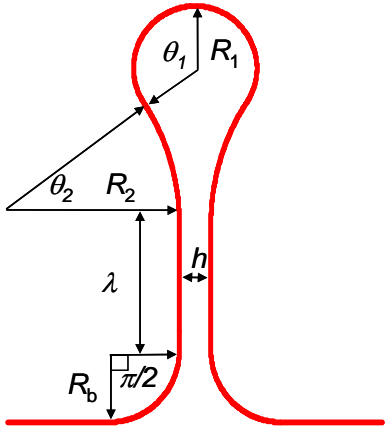

Fig. S2. Schematics of the *standing collapsed graphene wrinkle*. The minimum energy from Eq. (3) corresponds to the values of the variational parameters $R_b = \sqrt{2\pi/(2+\pi)}\,R_0$, $R_1 \approx 0.967 R_0$, $R_2 \approx 3.401 R_0$, $\theta_1 \approx 0.685\pi$, $\theta_2 \approx 0.185\pi$, where $R_0 = 6.8$ Å.

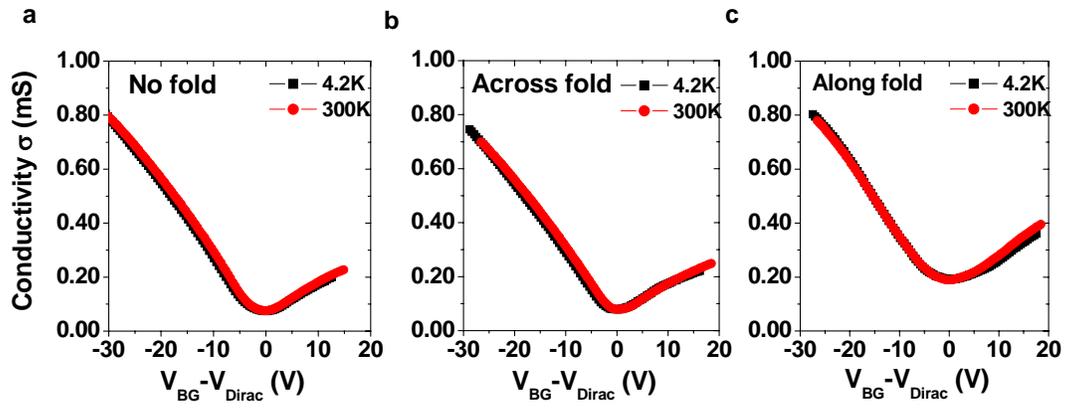

Fig.S3. Conductivity as a function of $V_{BG}$-$V_{Dirac}$ at 4.2K and 300K for a graphene device (a) with no fold, (b) measured across the fold and (c) along the fold.

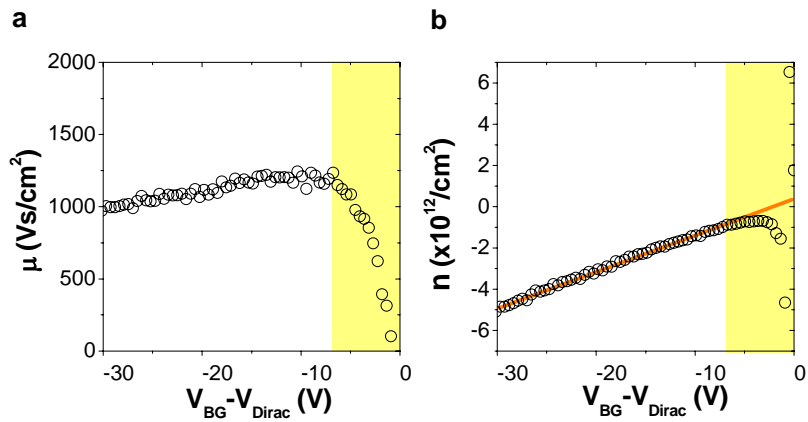

Fig.S4. Extracted (a) Hall mobility and (b) carrier density in graphene at 4K, via the standard Hall measurement procedure.

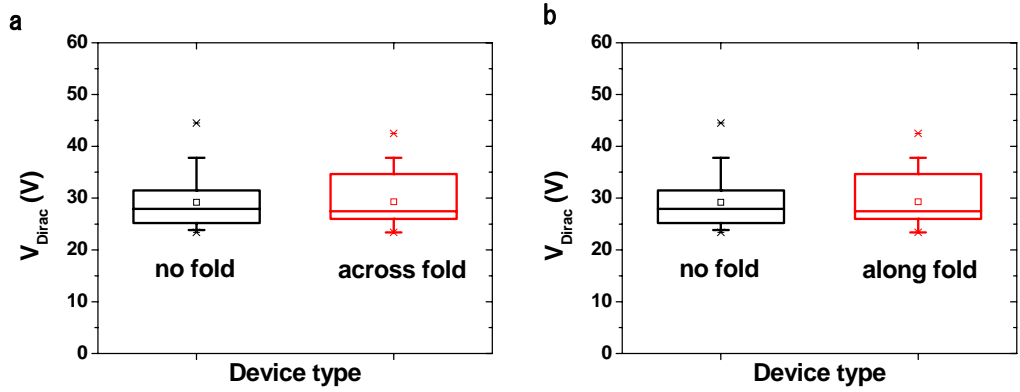

Fig. S5. Statistics of Dirac voltage of graphene Hall-bars (a) across fold vs no fold, and (b) along fold vs no fold. The statistics are based on the data from 42 devices.